\documentclass[sigconf]{acmart}

\usepackage{booktabs} 
\usepackage{balance}

\usepackage{color}
\newcommand{\bnote}[2]{
	\fbox{\bfseries\sffamily\scriptsize#1}
	{\sf\small$\blacktriangleright$\textit{#2}$\blacktriangleleft$}}

\newif\ifcomment
\commenttrue
\ifcomment
\newcommand{\elder}[1]{\bnote{Elder}{\textcolor[rgb]{1,0,1}{#1}}}
\newcommand{\marcelo}[1]{\bnote{Marcelo}{\textcolor{cyan}{#1}}}
\newcommand{\review}[1]{\bnote{review\textsubscript{3}}{\textcolor{red}{#1}}}
\newcommand{\reviewOne}[1]{\bnote{review\textsubscript{1}}{\textcolor{red}{#1}}}
\newcommand{\reviewTwo}[1]{\bnote{review\textsubscript{2}}{\textcolor{red}{#1}}}
\else
\newcommand{\elder}[1]{}
\newcommand{\marcelo}[1]{}
\newcommand{\review}[1]{}
\newcommand{\reviewOne}[1]{}
\newcommand{\reviewTwo}[1]{}
\fi

\newif\ifcolorred
\colorredfalse
\ifcolorred
\newcommand{\newtext}[1]{\textcolor{red}{#1}}
\else
\newcommand{\newtext}[1]{\textcolor{black}{#1}}
\fi

\usepackage{listings}
\usepackage{color}
\definecolor{dkgreen}{rgb}{0,0.6,0}
\definecolor{gray}{rgb}{0.5,0.5,0.5}
\definecolor{mauve}{rgb}{0.58,0,0.82}

\usepackage{caption}
\usepackage{hyperref}
\hypersetup{colorlinks=false, pdfborder={0 0 0}}
\usepackage{enumitem}
\usepackage{graphicx}
\usepackage{subfig}
\usepackage[]{natbib}
\usepackage{multirow}
\usepackage{xcolor}

\usepackage{caption}
\newcommand{\codigo}{\text{Code}}
\newfloat{JavaCode}{htbp}{loa}
\floatname{JavaCode}{\codigo}
\newcommand{\smell}[1]{\emph{#1}}

\setcopyright{none}

\acmYear{} 
\acmConference[SBES '21]{Brazilian Symposium on Software Engineering}{September 27-October 1, 2021}{Joinville, Brazil}
\acmBooktitle{Brazilian Symposium on Software Engineering (SBES '21), September 27-October 1, 2021, Joinville, Brazil}
\acmPrice{} 
\acmDOI{} 
\acmISBN{} 

\begin{document}
\title{On the Interplay of Smells \smell{Large Class}, \smell{Complex Class} and \smell{Duplicate Code}}



 \author{Elder Vicente de Paulo Sobrinho}
 \orcid{0000-0001-7735-6732}
 \affiliation{%
   \institution{Federal University of Tri\^angulo Mineiro}
   \streetaddress{Department of Electrical Engineering}
   \city{Uberaba}
   \state{MG}
   \country{Brazil}
 }
 \email{elder.sobrinho@uftm.edu.br}

 \author{Marcelo de Almeida Maia}
 \orcid{0000-0003-3578-1380}
 \affiliation{%
   \institution{Federal University of Uberl\^andia}
   \streetaddress{Faculty of Computing}
   \city{Uberl\^andia}
   \state{MG}
   \country{Brazil}
 }
 \email{marcelo.maia@ufu.br}

\renewcommand{\shortauthors}{Sobrinho E. V. P. and Maia M. A.}

\begin{abstract}
Bad smells have been defined to describe potential problems in code, possibly pointing out refactoring opportunities. Several empirical studies have highlighted that smells have a negative impact on comprehension and maintainability. Consequently, several approaches have been proposed to detect and restructure them. 
However,  studies on the inter-relationship of occurrence of different types of smells  in source code are still lacking, especially those focused on the quantification of this inter-relationship.
In this  work, we aim at understand and quantify the possible the inter-relation of smells \smell{Large Class - LC}, \smell{Complex Class - CC} and \smell{Duplicate Code - DC}. In particular, we investigate patterns of LC and CC regarding the presence or absence of duplicate code.
We conduct a quantitative study on five open source projects,  and also a  qualitative analysis to measure and understand the association of specific smells.
As one of the main results, we highlight that there are "occurrence patterns" among these smells, for example: either in \smell{Complex Class} or in the co-occurrence of \smell{Large Class} and \smell{Complex Class}, clones tend to be more prevalent in highly complex classes than less complex classes.
The found patterns could be used to improve the performance of detection tools or even help in refactoring tasks.

\end{abstract}


\maketitle

\section{Introduction}

Software systems need to evolve continuously to cope with new requirements and environment changes. High-quality source code plays an important role in this context because the code itself is required to be easy to understand, analyze, change, maintain, and reuse \citep{Jiang2014}. However, software developers eventually produce sub-optimal code (not at the highest standard), possibly introducing design problems, i.e., they produce code structures that violate fundamental principles in software engineering, such as, high cohesion and low coupling.
Bad smells have been proposed as a metaphor for sub-optimal code structures, and have gained attention after \citet{fowler1999refactoring} proposing that those structures  could be refactored in a systematic way to produce better quality code.
On the other hand, bad smells may not be as harmful as generally claimed, in other words, they are not always associated with undesirable or problematic situations \cite{Sobrinho2018TSE}. For instance, \citet{P266} report that bugs are not significantly associated with \smell{Duplicated Code}. Also, in some situations, writing code with the presence of bad smells is even the best option for developers \citep{P2}.  


Despite the large body of knowledge already produced on code bad smells, there is still room for investigating better this topic \cite{Sobrinho2018TSE}. There may be common sense knowledge that may not hold as expected in real-world projects. An interesting example has been shown by \citet{TSE2017Tufano}, where their findings \textit{``contradict common wisdom, showing that most of the smell instances are introduced when an artifact is created and not as a result of its evolution"}. 
In our work, we start from the observation that smells seems to be a fragmented metaphor for analyzing code entities because they are defined to characterize a single sub-optimal structure in the code. 
However, code entities do not manifest only a pure view of such metaphors, i.e., an entity may manifest more than one kind of smell simultaneously. So, in this work,  we target a phenomenon that is not much addressed in the literature: the co-occurrence of different smell types in the same code entity.
We aim at investigating the inter-relation between multiple kinds of bad smells. In particular, we study the inter-relation of smells \smell{Duplicate Code} (DC), \smell{Large Class} (LC) and \smell{Complex Class} (CC). 
For instance, we aim at verifying the actual extent of common wisdom raised by \citet{fowler1999refactoring}, when they explain the smell \smell{Large Class} and suggest an interplay with \smell{Duplicate Code}: ``\textit{...a class with too much code is prime breeding ground for duplicated code...}".
Moreover, there could be other scenarios, for instance, where complex and large classes could contain code snippets  cloned in  other also complex and large classes.  This potential pattern could be possibly useful for software engineers, in particular, to improve their rules of code review (e.g., check the possible existence of clones in complex and large classes to improve three kinds of sub-optimal code structures simultaneously). Summing up, we expect that such possible inter-relationships could be unveiled by analyzing how the co-occurrence of complex and large class smells may actually affect the prevalence of clones.

The smells \smell{Duplicate Code}, \smell{Large Class} and \smell{Complex Class} are the focus of this paper because: i) these smells are recurrent in literature but, to the best of our knowledge,  no paper has investigated their inter-relation; ii) those smells  are commonly found in the source code of several projects and the number of instances of them is large enough to allow statistical analysis; iii) semantically there seems to be a relationship between themselves, so 
we aim to investigate if the following  hypothesis actually holds:
the complex entities with many control flow statements could be prone to the occurrence of clones, in particular, when a large number of conditional expressions are present and perform the same code or slightly different codes (differing only in their conditions)\footnote{\url{https://refactoring.guru/smells/duplicate-code}}. 
Moreover, there is still the fact that the code smell metaphor has been proposed to deliver a boolean value, e.g., a class is large or not. However, the intensity of a smell may indicate its severity. For instance, \citet{icsme2016teenspirit} have improved a bug prediction model adding the intensity of the smell as a predictor. In this paper, we study the interplay of smells according to their co-occurrence and intensity in different combinations. 
The main contributions of the paper are: i) we expand the knowledge about the  prevalence of the co-occurrence of \smell{Duplicate Code}, \smell{Large Class} and \smell{Complex Class}.  In particular, we investigate patterns of \smell{LC} and \smell{CC} regarding the presence or absence of duplicate code; ii) we reveal the occurrence of patterns among smells, e.g., the prevalence of clones is much associated to the occurrence of the smell \smell{Complex Class} than to the isolated occurrences of \smell{Large Class}; iii) to the best of our knowledge, this is the first work to use intensity of smells to finding patterns among several smells; iv) we also introduce practical implications of our results on strategies used to detect smells and on to refactoring planning.

\textbf{Structure.} The remainder of this paper is structured as follows: in the next section, we present  background information 
and review the related work. The Section \ref{StudySetting} describes the design of our empirical study, while Section \ref{studyresults} reports the obtained results. Next, the Section \ref{Discussion} discusses and provides a qualitative perspective of our results. In Section \ref{ThreatsValidity}, we present the limitations and threats. Finally, in Section  \ref{Conclusion}, our conclusions are drawn. 

\section{Background and Related Work}  \label{RelatedWork}

Next, we present some concepts used 
and discuss related work. 

\subsection{Background}
We use the concept of  interrelationship  and intensity of smells. First, we show different types of interrelationships, and then we define a metric to measure how intense each smell instance is.

\subsubsection{Interrelationship of smells} ~

Smells could be interrelated in several different manners. We present three possible different  forms of smell interrelationship: \textit{by co-occurrence, by static dependency} and \textit{by commit dependency}. In our work, we consider only the interrelationship by co-occurrence.

\begin{figure}[H]
	\includegraphics[width=0.85\linewidth]{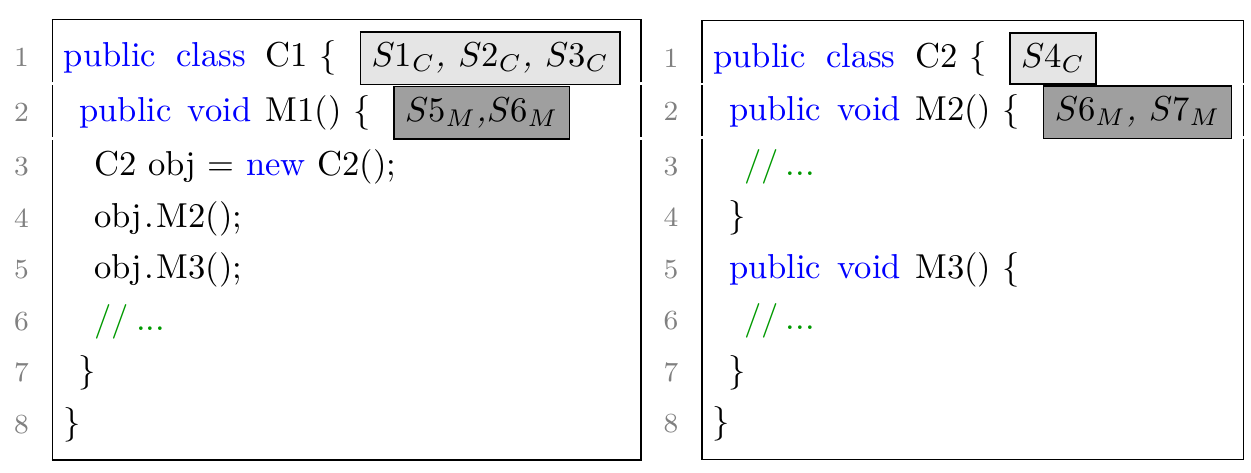}
	\caption{\footnotesize Interrelationship of smells (co-occurrence \& dependency).}
	\label{InterrelationshipOfSmells}
\end{figure}

The class \textit{C1} (\figurename~\ref{InterrelationshipOfSmells}) has three different kinds of smells (\textit{$S1_C$, $S2_C$, $S3_C$}) and the class \textit{C2} show another kind of smell (\textit{$S4_C$}). These two entities have four smells and they occur at the level of class. In the method level, these classes have other three kinds of smells (\textit{$S5_M$, $S6_M$, $S7_M$}). Observe that the method \textit{M1} and \textit{M2} has the same kind of smell (\textit{$S6_M$}). 

Smells can be interrelated by the source code structure, either by a co-occurrence or by a static dependency. The interrelationship by \textit{co-occurrence} considers the existence of smells only in one entity at a time, e.g., the file of the class \textit{C1} have five smells (\textit{$S1_C$, $S2_C$, $S3_C$, $S5_M$, $S6_M$}) and they can be grouped by the granularity of entity (class/method). The interrelationship by \textit{static dependency} takes into account the coupling with other entities and the existence of smells in them, e.g., the method \textit{M1} inside the class \textit{C1} invokes the methods \textit{M2} and \textit{M3} of the class \textit{C2}. Thus, these seven smells (\textit{$S1_C$, $S2_C$, $S3_C$, $S4_C$, $S5_M$, $S6_M$, $S7_M$}) are interrelated by  a \textit{static dependency} of the respective entities, and they can be grouped by granularity of each entity.
The change history of source code could also be used to interrelate  smells, in particular, using \textit{commits}. Considering the methods \textit{M2} and \textit{M3}, there is no interrelationship by the source code structure. Nevertheless, taking into account  \textit{commits}, if  those methods were possibly changed inside the same \textit{commit}, they would be commit-interrelated, meaning that possibly  smells $S6_M$, $S7_M$ would have some impact on \textit{M3}. 

Considering that the smells can be interrelated in different forms, 
we clarify that in this paper only the interrelationship of smells by \textit{co-occurrence} at the class level are studied.

\subsubsection{Intensity of Smells} ~

 \citet{P284} describes an approach to compute smell intensity based on refactoring techniques (\textit{pull-up method, extract method}) used to remove the smells. They compute the smell intensity, considering the size (LOC) and the number of parameters necessary to complete the refactoring task. \citet{PA406} consider the threshold of metrics used on smell detection to classify the level of intensity of smells. On their approach, the more the value of a given metric exceeds your threshold value, the greater is the smell intensity. Based on the distribution of these metrics, they classified the smell intensity in five range (\textit{Very-Low, Low, Mean, High, Very-High}).

We also consider the smell intensity on this study of interrelationship of smells. The intensity of smells could possibly improve the identification of patterns, e.g., as the smells become more critical, they could possibly end up associating themselves with other types of smells. Thus, in context of smell \smell{Complex Class}, we can analyze whether when  complexity increases, the prevalence of other types of smells also increases. 

We use an approach similar that proposed by \citet{PA406} to compute the intensity of smells, but we simplify on only two ranges (\textit{Low} or \textit{High}). 
We decided not to use a continuous variable for intensity because it would be more difficult to find any pattern using such representation, so a binary representation would have more chances to explain possible patterns.
Let $E=\{e_{1_{S_{1}}},~e_{2_{S_{2}}},...,~e_{n_{S_{n}}}\}$ be a set of distinct entities $e_{i_{S_{i}}}$, where in each one co-occurs a set $S_{i}$  of different smells ($1\leq i \leq n$). Let $S = \bigcup_{i=1}^{\;n} S_{i}$ be the set of all distinct smells. Let $s_j$ be a smell in $S$. In  order to compute the intensity of a smell $s_j$ in an entity $e_i$, we extract a subset $E_{s_{j}}$ of $E$ with all entities of $E$ occurring $s_j$, i.e., $E_{s_{j}}=\{e_{S'} \mid e_{S'} \in E  \land s_{j} \in S' \}$. Then, let the set $M_{j}$ contain the computed metrics\footnote{ \smell{Complex Class} = \textit{Cyclomatic Complexity}, \smell{Large Class} = Number of methods declared + Number of attributes declared, both used by DECOR. \smell{Duplicate Code} = number of tokens, used by PMD tool.} 
$m_{j}(e)$ used to identify this smell $s_j$ in each $e \in E_{s_{j}}$. 
Finally, for each entity $e_i$ ($1\leq i \leq n$) with smell $s_j$ ($1 \leq j \leq S$), we use the Equation \ref{EqIntensity} to classify their intensities.

\begin{equation}
Intensity_{s_{j}}(e_{i}) =\left\{
\begin{matrix}
High, & if~m_{j}(e_i) \ge median(M_{j})\\
Low,  & otherwise 
\end{matrix}\right.
\label{EqIntensity}
\end{equation}

\subsection{Related Work}

According to \citet{7886924} each smell alone may represent only a partial embodiment of a design problem. They suggest that smells tend to "flock together" to realize a design problem. Thus, they investigated whether and how smelly code relationships can help developers to locate design problems. To achieve this purpose, they propose a strategy to identify groups of inter-related smells. This strategy is composed of  syntactic and semantic forms used to connect elements of a software, as example: two smells are syntactically related if their host program elements are connected through method calls or inheritance relationships. Their analysis indicates that certain forms to connect elements are consistent indicators of both congenital and evolutionary design problems, with accuracy often higher than 80\%. They also found the combined use of syntactic and semantic forms to connect elements of a software represents a more effective approach for locating design problems.

Recent studies suggest that developers should ignore smells occurring in isolation. 
Instead, they should focus on analyzing smell agglomerations, e.g., a entity affected by multiple smells. However, there is limited understanding if developers can effectively identify a design problem in stinkier code. Developers may struggle to make a meaning out of inter-related smells affecting the same program location. In this context, \citet{Oizumi:2017:RDP} applied an approach to analyze if and how developers can effectively find design problems when a program location is affected by multiple smells. The analysis revealed that only 36.36\% of the developers found more design problems when they explicitly use smell agglomerations to identify design problems as compared to single occurrence of smells. On the other hand, 63.63\% of the developers reported much lesser false positives. Developers reported that analyses of smell agglomerations scattered in class hierarchies or packages are often difficult, time consuming, and requires proper visualization support. Moreover, it remains time-consuming to discard stinky program locations that do not represent design problems.

\citet{SAC2021Figueiredo} have also characterized co-occurrences as agglomerations, i.e., when two or more bad smells occur in the same snippet of code. They studied four kinds of bad smells: Large Class, Long Method, Feature Envy and Refused Bequest using association rules and found that classes with two or more smells are frequent in the source code, even when the smells in a class are of the same type. They also found that agglomerations are highly spread in the source code having significant effect on modularity metrics.

\citet{Fontana:2015} focus their attention on the possible relations existing among code  smells and their co-occurrence (how many entities are affected by more than one smell), with the aim to find and detect some architecturally relevant code smells. They found that a significant percentage of the detected instances has a relation with other instances. For example, 26\% of \smell{God Classes} use at least a \smell{Data Class}, and that 53\% and 70\% of methods affected by \smell{Shotgun Surgery} and \smell{Dispersed Coupling}, respectively, are called from (at least) one class or method affected by a code smell. This observation confirms other results and theories proposed in different studies, suggesting that code smells tend to cluster together and interact in many ways, and that clusters of smells have more effect on maintainability than isolated smells. 

According to \citet{7882010}, there is little knowledge regarding which smell types tend to co-occur in  code. To enlarge the knowledge on the phenomenon, they provide a large-scale replication of previous studies, taking into account 13 smell types on a dataset composed of 395 releases of 30 software systems. They identified six pairs of code smells that co-occur very often, some of these co-occurrences are quite expected (e.g., \smell{Long Method} and \smell{Spaghetti Code}), others are not (e.g., \smell{Message Chains} and \smell{Refused Bequest}), recalling the need for studying more deeply the reasons behind their appearance and their apparent relationships.

\citet{8009932} investigated the relationship between code clones and 15 anti-patterns (smells) documented by \citet{Brown:1998:ARS:280487}. They conducted the study on three open-source  systems and  results show that clones and anti-patterns is a frequent observation: at least, more than 52\% of anti-pattern classes have clones, while 59\% to 78\% of classes with clones are participating in anti-patterns. They also observe that classes having clones and anti-patterns are significantly more fault-prone than other classes. The analysis also reveals that a high number of smell co-occurrence among classes increases the risk of fault induction in software systems and decrease their reliability. Finally, they suggest that the detection of smell co-occurrence among classes helps to manage change commits and to avoid faults induction during the maintenance activities.

\citet{6606614}  investigated interactions (e.g., smells that were co-located in the same artifact) among twelve different smells and how these interactions can lead to maintenance problems. Analyzing how  developers conducted tasks on four different systems, they found  evidence that certain inter-smell relations were associated with problems during maintenance, e.g.: i) artifacts containing \smell{ISP Violation} relates to the presence of \textit{inconsistent design}; ii) classes contained many methods that accessed data/methods from different areas of the system (i.e., methods displaying \smell{Feature Envy}) are related to the faults because developers missed areas of the code that needed to be consistently changed after changes were done on the methods displaying \smell{Feature Envy}. 

Previous studies have not investigated the relation between the smells \smell{Duplicate Code}, \smell{Large Class} and \smell{Complex Class}. Moreover,  studies that take into account the intensity of these smells are also lacking. 
Thus, in this study we sought to provide additional evidence on the relationship between some kinds of smells.

\section{Study Setting} \label{StudySetting}
In this section we define the research questions of the study, explaining the process followed to assess them.

\subsection{Research Questions} \label{ResearchQuestions}

The \textit{goal} of the study is to analyze the interplay of co-occurrence of smells \smell{Duplicate Code (DC)}, \smell{Large Class (LC)} and \smell{Complex Class (CC)}. The motivation for this analysis is to assess to which extent the several combinations of smells \smell{LC} and \smell{CC} in different combinations of (co-)occurrence associates with cloning in the respective entities. To organize the analysis, the study has been divided into two parts: i) analysis considering the absolute frequency of smells; ii) analysis considering the intensity of smells.


In the first part of our research questions, we investigate what happens with the prevalence of clones as the frequency of \textit{CC} and \textit{LC} smells co-occurring in a class changes. 

\begin{figure}[tbh]
	\includegraphics[width=0.85\linewidth]{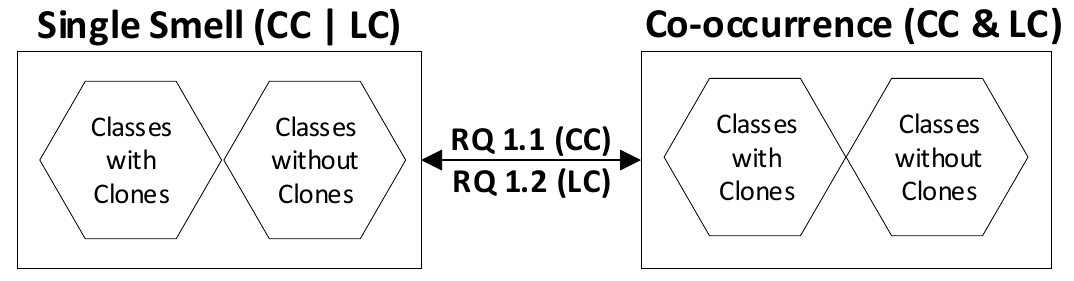}
	\caption{Diagram of \ref{RQ-CC_LCCC} and \ref{RQ-LC_LCCC}.}
	\label{FigRQ1_x}
\end{figure}

\begin{enumerate}[label*={\textbf{RQ1.\arabic*}},leftmargin=1.25cm, ref={RQ1.\arabic*}] 		
	\global\def\RQCCLCCC{Is there a difference in the prevalence of clones comparing the entities where only \textit{CC} occur and those entities where \textit{CC} and \textit{LC}  co-occur? If so, how much this difference is associated to the smells \textit{CC} and/or \textit{LC}? }

 	\item \label{RQ-CC_LCCC} \RQCCLCCC
	
	\item[\textbf{}] \newtext{As shown in  \figurename~\ref{FigRQ1_x}, we compare two groups of classes: 1) on the left, all classes with smells \smell{CC}, not occurring the smell \smell{LC} and 2) on the right, all classes with co-occurrence of smells \smell{CC \& LC}. In both groups, we have classes with and without the smell \smell{DC}.}



	\global\def\RQLCLCCC{Is there a difference in the prevalence of clones comparing the entities where \textit{LC} and \textit{CC} co-occur and those entities where only \textit{LC} occur? If so, how much this behavior is associated to the smells \textit{LC} and/or \textit{CC}? }
	
	\item \label{RQ-LC_LCCC} \RQLCLCCC
	
	\item[\textbf{}] \newtext{Similarly to the previous RQ as also shown in \figurename~\ref{FigRQ1_x}, we also compare two groups of classes: 1) on the left, all classes with smells \textit{LC} and that does not have the smell \textit{CC} and 2) on the right, all classes with co-occurrence of smells \textit{CC} \& \textit{LC}. Again, both groups have classes with and without the smell \textit{DC}.}
	
\end{enumerate}

In the second part of our research questions, we evaluate what happens to the clone prevalence when the intensity of a smell or a set of smells becomes more critical. 

\begin{figure}[tbh]
	\includegraphics[width=0.85\linewidth]{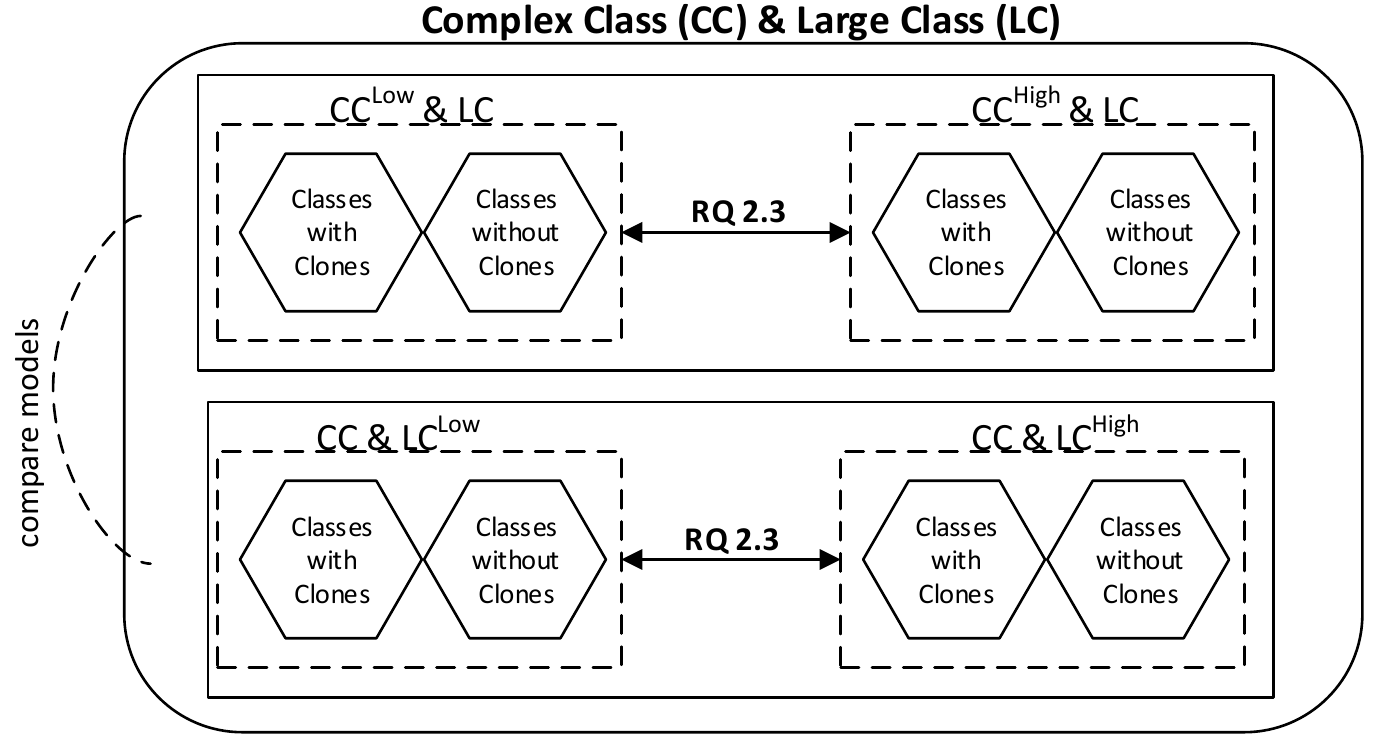}
	\caption{Diagram of \ref{RQ-RQ_LCCCIntensity_Clone}.}
	\label{FigRQ2.3}
\end{figure}

\begin{enumerate}[label*={\textbf{RQ2.\arabic*}},leftmargin=1.25cm, ref={RQ2.\arabic*}] 		
	\global\def\RQCCIntensityClone{Is there any association between the prevalence of clones and the intensity of smell \smell{Complex Class}?}
	
	\item \label{RQ-CCIntensity_Clone} \RQCCIntensityClone
	
	\item[\textbf{}] This research question investigates the relationship between the occurrence of clones and entities classified as less complex and those more complex. \newtext{\figurename~\ref{FigRQ2_x} shows that, for RQ2.1, we compare two groups of classes, 1) on the left, all classes with only smell CC in \emph{Low intensity} and 2) on the right, all classes with this smell in \emph{High intensity}. In both groups, we have classes with and without smell DC. Additionally, in these RQs, we do not consider the co-occurrence of smells (CC \& LC).}


	\global\def\RQLCIntensityClone{Is there any association between the prevalence of clone and the intensity of smell \smell{Large Class}?}

	\item \label{RQ-LCIntensity_Clone} \RQLCIntensityClone
	
	\item[\textbf{}] Similarly to the previous RQ2.1, but now for \textit{LC}, this  question investigates the relationship between the occurrence of clones and entities classified as \smell{Large Class} only. This relationship is compared under two levels of \smell{Large Class} intensity (\smell{\textit{Low}} and \smell{\textit{High}}).

	\begin{figure}[tbh]
	\includegraphics[width=0.85\linewidth]{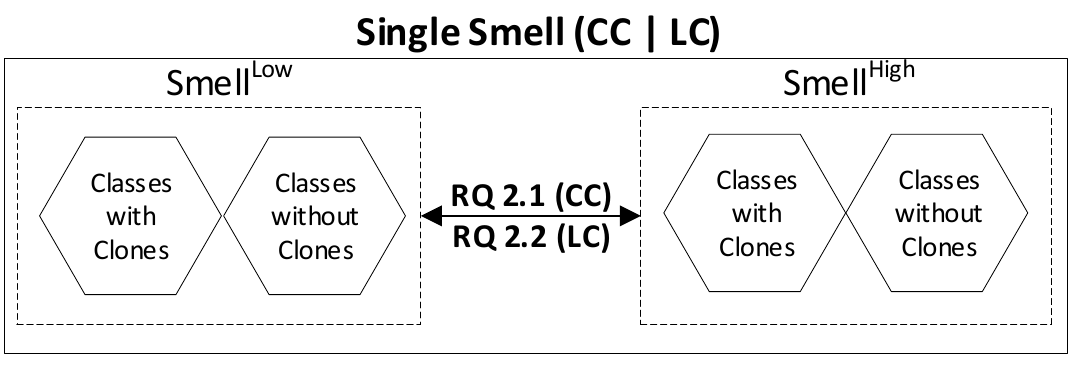}
	\caption{Diagram of \ref{RQ-CCIntensity_Clone} and \ref{RQ-LCIntensity_Clone}.}
	\label{FigRQ2_x}
\end{figure}

	\global\def\RQLCCCIntensityClone{Is there any association between the prevalence of clone and the intensity and co-occurrence of smells \smell{Large Class} and \smell{Complex Class}?}

	\item \label{RQ-RQ_LCCCIntensity_Clone} \RQLCCCIntensityClone
	
	\item[\textbf{}] \newtext{This research question investigates the relationship between the occurrence of clones and co-occurrence of smells \smell{Large Class} and \smell{Complex Class}, which are analyzed 
	in two levels of intensity (\emph{Low} vs. \emph{High}).
	As shown in \figurename~\ref{FigRQ2.3}, we formulated two models based on smell co-occurrence and intensity to analyze the impact of the intensity of  \smell{Large Class} and \smell{Complex Class}: \smell{$CC^{Low}~\&~LC$} vs \smell{$CC^{High}~\&~LC$} and \smell{$CC~\&~LC^{Low}$} vs \smell{$CC~\&~LC^{High}$}.
	 }

\end{enumerate}

\subsection{Study Variables} \label{Variables}

The \textbf{dependent variables} considered in our study, for all the research questions, are the presence or absence of clones being observed across different software. In other words, this variable is dichotomous ($Y_{DC}$) and denotes the presence of smell \smell{Duplicate Code}. The \textbf{independent variables} are the factors related to the smell intensity (LC,CC), single occurrence of smell (LC,CC) and/or co-occurrence of smells (LC,CC) and are defined according to the research questions:

	\textbf{\ref{RQ-CC_LCCC}}. We have a categorical variable with two possible values:  one denoting the  occurrence of the smell \smell{Complex Class} only 
	and another denoting the co-occurrence of smells \smell{Complex Class} and \smell{Large Class}. 
	The model is:  
	$Y_{DC} = \beta_0 + \beta_1 X_{Single \_ Co\mbox{-}occurrence}$ 
	
	\textbf{\ref{RQ-LC_LCCC}}. Similar to the previous question, we also have a categorical variable limited to two values: one denoting the  occurrence of the smell \smell{Large Class} only 
	and another denoting the co-occurrence of smells \smell{Complex Class} and \smell{Large Class}.. Thus, the model also is similar.
	
	\textbf{\ref{RQ-CCIntensity_Clone}}. For this research question, the categorical variable denotes the intensity of the smell \smell{Complex Class} in two levels ($Low | High$). The model may be represented as: $Y_{DC} = \beta_0 + \beta_1 X_{Low \_ High}$.
	
	\textbf{\ref{RQ-LCIntensity_Clone}}. In this case, the categorical variable $X$ denotes the  intensity of the smell \smell{Large Class}. Thus, the model is similar to the previous question.
	
	\textbf{\ref{RQ-RQ_LCCCIntensity_Clone}}. The categorical variable denotes the intensity ($Low | High$) of two smells (\smell{Complex Class} and \smell{Large Class}). In particular, the intensity of these smells are investigated when they co-occur in the same class. Thus, we have two models: i) considering the intensity of the smell \smell{Large Class} ($Y^{\prime}_{DC} = \beta_0 + \beta_1 X_{CC\&LC_{Low \_ High}}$), and ii) based on the intensity of the smell \smell{Complex Class} ($Y^{\prime\prime}_{DC} = \beta_0 + \beta_1 X_{LC\&CC_{Low \_ High}}$). 

\subsection{Studied Systems and Data Extraction}
The study consists of 5 Java open source software systems and having different sizes and domains (see  \tablename~\ref{ProjectsAnalyzed}). 
ArgoUML is a UML diagramming system in Java. 
Lucene is a a high-performance, full-featured text search engine library. Cassandra is a database management system. Hadoop is a tool for distributed computing. Apache Ant is a build tool and library specifically conceived for Java applications (though it can be used for other purposes). 

\begin{table}
	\scriptsize
	\centering
	\def\arraystretch{1.4}
	\setlength{\tabcolsep}{1.2pt}
	\caption{Studied open source projects.}
	\label{ProjectsAnalyzed}
	\begin{tabular}{ccccccccc}
		\hline
		
		\multicolumn{1}{c}{\multirow{2}{*}{\textbf{Project}}} &
		\multicolumn{1}{c}{\multirow{2}{*}{\textbf{Version}}} &
		\multicolumn{1}{c}{\multirow{2}{*}{\textbf{\shortstack{Java\\ File}}}} &
		\multicolumn{1}{c}{\multirow{2}{*}{\textbf{LOC}}} &
		\multicolumn{1}{c}{\multirow{2}{*}{\textbf{\shortstack{Snippets\\ Cloned}}}} & 
		\multicolumn{1}{c}{\multirow{2}{*}{\textbf{\shortstack{Classes \\With \\ Clones}}}} &
		\multicolumn{1}{c}{\multirow{2}{*}{\textbf{\shortstack{\textsc{Large}\\ \textsc{Class} }}}} &
		\multicolumn{1}{c}{\multirow{2}{*}{\textbf{\shortstack{\textsc{Complex}\\ \textsc{Class} }}}} &
		\multicolumn{1}{c}{\multirow{2}{*}{\textbf{\shortstack{Co-occurrence \\ LC \& CC }}}}
		\\
		&
		&
		&
		&
		&
		&
		&
		&
		\\
		\hline
		
		\multicolumn{1}{c}{\multirow{1}{*}{ArgoUML}} &
		\multicolumn{1}{c}{\multirow{1}{*}{0.34}} &
		\multicolumn{1}{c}{\multirow{1}{*}{1233}} &
		\multicolumn{1}{c}{\multirow{1}{*}{105795}} &
		\multicolumn{1}{c}{\multirow{1}{*}{214}} &
		\multicolumn{1}{c}{\multirow{1}{*}{79}} &
		\multicolumn{1}{c}{\multirow{1}{*}{63}} &
		\multicolumn{1}{c}{\multirow{1}{*}{37}} &
		\multicolumn{1}{c}{\multirow{1}{*}{119}} 
		\\

		\multicolumn{1}{c}{\multirow{1}{*}{Cassandra}} &
		\multicolumn{1}{c}{\multirow{1}{*}{3.11}} &
		\multicolumn{1}{c}{\multirow{1}{*}{2062}} &
		\multicolumn{1}{c}{\multirow{1}{*}{333211}} &
		\multicolumn{1}{c}{\multirow{1}{*}{760}} &
		\multicolumn{1}{c}{\multirow{1}{*}{31}} &
		\multicolumn{1}{c}{\multirow{1}{*}{38}} &
		\multicolumn{1}{c}{\multirow{1}{*}{44}} &
		\multicolumn{1}{c}{\multirow{1}{*}{110}} 
		\\
		
		\multicolumn{1}{c}{\multirow{1}{*}{Lucene}} &
		\multicolumn{1}{c}{\multirow{1}{*}{6.2.1}} &
		\multicolumn{1}{c}{\multirow{1}{*}{3848}} &
		\multicolumn{1}{c}{\multirow{1}{*}{533926}} &
		\multicolumn{1}{c}{\multirow{1}{*}{2255}} &
		\multicolumn{1}{c}{\multirow{1}{*}{137}} &
		\multicolumn{1}{c}{\multirow{1}{*}{54}} &
		\multicolumn{1}{c}{\multirow{1}{*}{100}} &
		\multicolumn{1}{c}{\multirow{1}{*}{160}} 
		\\
		
		\multicolumn{1}{c}{\multirow{1}{*}{Hadoop}} &
		\multicolumn{1}{c}{\multirow{1}{*}{2.6.0}} &
		\multicolumn{1}{c}{\multirow{1}{*}{1417}} &
		\multicolumn{1}{c}{\multirow{1}{*}{194518}} &
		\multicolumn{1}{c}{\multirow{1}{*}{605}} &
		\multicolumn{1}{c}{\multirow{1}{*}{25}} &
		\multicolumn{1}{c}{\multirow{1}{*}{26}} &
		\multicolumn{1}{c}{\multirow{1}{*}{25}} &
		\multicolumn{1}{c}{\multirow{1}{*}{68}} 
		\\
		
		\multicolumn{1}{c}{\multirow{1}{*}{Ant}} &
		\multicolumn{1}{c}{\multirow{1}{*}{1.8.2}} &
		\multicolumn{1}{c}{\multirow{1}{*}{1182}} &
		\multicolumn{1}{c}{\multirow{1}{*}{127042}} &
		\multicolumn{1}{c}{\multirow{1}{*}{170}} &
		\multicolumn{1}{c}{\multirow{1}{*}{33}} &
		\multicolumn{1}{c}{\multirow{1}{*}{10}} &
		\multicolumn{1}{c}{\multirow{1}{*}{33}} &
		\multicolumn{1}{c}{\multirow{1}{*}{96}} 
		\\ 
		
		\hline		
	\end{tabular}
\end{table}

To answer our research questions, we first need to detect the smells of the studied systems. To this aim we use existing tools, PMD\footnote{\url{https://pmd.github.io/}} and DECOR\footnote{\textit{DEtection \& CORrection} --- \url{https://bitbucket.org/ptidejteam/}}. 
The first tool is a token-based approach used to detect \smell{Duplicate Code}, in particular the type I\footnote{\textit{Code are identical except for variations in whitespace, layout and comments} \cite{P271}.}.
The output of PMD denotes which snippets of a class also exists in other parts of the system. This output enables us to classify each clone according to their locality: i) \textbf{intra-class}, the clones of a class occur only inside the class itself; ii) \textbf{inter-classes}, a snippet of code classified as clone occurs only between different classes; iii) \textbf{mix-classes},  refers to clones occurring inter- and intra-class simultaneously.
We use this tool because it enables us to classify the locality of clones and because \citet{Roy2009a} reports an extensive comparison of clone detection tools and they reported that PMD is good at detecting identical clones. 
Moreover, the tool DECOR identifies smells using detection rules based on the values of internal quality metrics\footnote{An example of detection rule exploited to identify \smell{Blob} classes can be found at \citep{P154}.}. In our paper, this tool is used identify the classes with the smells \smell{Large Class} and \smell{Complex Class}. The choice of using DECOR is driven by the fact that (i) it is a state-of-the-art smell detector having a high accuracy in detecting smells \citep{P154}; and (ii) it applies simple detection rules that allow it to be very efficient.


Thus, for each class of the analyzed systems, we identify whether the smells \smell{Large Class}, \smell{Complex Class} and/or \smell{Duplicate Code} are present. For smelly classes, we also collect the numerical values of metrics used to identify each smell, e.g., for \smell{Complex Class} we collect the \textit{Cyclomatic Complexity} (McCabe) and for \smell{Large Class} the metric NMD (number of methods declared) and NAD (Number of attributes declared), and for \smell{Duplicate Code} the number of tokens of clone. These values are used to classify the intensity of smells (see Equation \ref{EqIntensity}).

\subsection{Analysis Method} \label{AnalysisMethod}
We build, for each object system and for each research question, logistic regression models \cite{hosmer2013applied} relating a (dichotomous) dependent variable - indicating the presence or absence of clones - with independent variables represented by the quality indicators (code smells and their intensities). 



Logistic regression assumptions relate to 
linearity, independence of errors, multicollinearity that are assumptions to consider.
Considering all models of our RQs, linearity is not violated, because they use categorical predictors \citep{field2012discovering}.
Considering that the classes analyzed in our study are independent and unrelated, we can assume the independence of errors.
Multicollinearity is not violated, because the models use only one categorical predictor. 
Regarding the data, for any pair of categorical variables, it is useful to set up a contingency table to show the cell frequencies. 
We need to check if there are not more than 20\% of cells with frequencies less than five, and any of the  frequencies equals zero. \citep{field2012discovering}.


For each model we analyze the Odds Ratio (OR) (\citet{schumacker2014learning}) which is given by $e^{\beta_i}$. The higher the OR for an independent variable, the higher its ability to explain the dependent variable. The interpretation of the OR for the model built using quality metrics (smells), changes between the kinds of models due to the different aims of each research question. In general, for our research questions, the OR for an independent variable indicates the increment of chances for a class to be subject of clones in consequent of a one-unit increase of the independent variable.

\section{Empirical study results} \label{studyresults}
This section discusses the results of our study, aimed at addressing the research questions. 
As explained, we performed a analysis of the assumptions of each logistic regression model. The underlying data used in this study is open and available \cite{elder_v_p_sobrinho_2021}.

\global\def\TitleRQOneX{The Number of Smells as a Factor}
\subsection{\TitleRQOneX}

This part of our study is related to the research questions that investigate the prevalence of clones in the (co-)occurrence of the specific smells \smell{Large} and \smell{Complex Class}, not taking the intensity of these smells into account.

\mbox{}\\
\noindent{\it \textbf{\ref{RQ-CC_LCCC}:} \RQCCLCCC~}
\begin{table}[H]
   
	\scriptsize
	\centering
	\def\arraystretch{2}
	\setlength{\tabcolsep}{2.2pt}
	\caption{Logistic Regression - \ref{RQ-CC_LCCC}.}
	\label{RQCCLCCC}
	\begin{tabular}{ccccccccc}
		\hline
		
		\multicolumn{1}{c}{\multirow{3}{*}{\textbf{Project}}} &
		\multicolumn{1}{c}{\multirow{3}{*}{\textbf{ \shortstack{$\beta_0$ \\ (SE)} }}} &
		\multicolumn{1}{c}{\multirow{3}{*}{\textbf{\shortstack{$\beta_1$ \\ (SE)}}}} &
		\multicolumn{1}{c}{\multirow{3}{*}{\textbf{AIC}}} &
		\multicolumn{1}{c}{\multirow{3}{*}{\textbf{R\textsuperscript{2}}}} &
		\multicolumn{3}{c}{\textbf{IC for $\beta_1$}} &
		\multicolumn{1}{c}{\multirow{3}{*}{\textbf{\shortstack{\textit{Significance} \\ \textit{of} \\ \textit{Predictor}}}}} \\ \cline{6-8} 
		
		\multicolumn{1}{c}{}&
		\multicolumn{1}{c}{} &
		\multicolumn{1}{c}{} &
		\multicolumn{1}{c}{}&
		\multicolumn{1}{c}{}&
		\multicolumn{1}{c}{\multirow{2}{*}{\textbf{\shortstack{Lower \\ (5\%)}}}} &
		\multicolumn{1}{c}{\multirow{2}{*}{\textbf{OR}}} &
		\multicolumn{1}{c}{\multirow{2}{*}{\textbf{\shortstack{Upper \\ (95\%)}}}} &
		\multicolumn{1}{c}{} \\  
		
		\multicolumn{1}{c}{}&
		\multicolumn{1}{c}{} &
		\multicolumn{1}{c}{} &
		\multicolumn{1}{c}{}&
		\multicolumn{1}{c}{}&
		\multicolumn{1}{c}{} &
		\multicolumn{1}{c}{} &
		\multicolumn{1}{c}{} &
		\multicolumn{1}{c}{} \\[-2.5ex] \hline
		
		
		\multicolumn{1}{c}{ArgoUML} &
		\multicolumn{1}{c}{\shortstack{-0.383\textsuperscript{$\blacklozenge$} \\ (0.33)}} &
		\multicolumn{1}{c}{\shortstack{-0.297\textsuperscript{$\blacklozenge$} \\ (0.38)}} &
		\multicolumn{1}{c}{205.91} &
		\multicolumn{1}{c}{0.003} &
		\multicolumn{1}{c}{0.394} &
		\multicolumn{1}{c}{0.742} &
		\multicolumn{1}{c}{1.415} &
		\multicolumn{1}{c}{$\chi^{2}$(1)=0.59, $p$=0.444} \\

		\multicolumn{1}{c}{Cassandra} &
		\multicolumn{1}{c}{\shortstack{-1.504\textsuperscript{$\triangledown$} \\ (0.39)}} &
		\multicolumn{1}{c}{\shortstack{-0.127\textsuperscript{$\blacklozenge$} \\ (0.46)}} &
		\multicolumn{1}{c}{143.77} &
		\multicolumn{1}{c}{0.001} &
		\multicolumn{1}{c}{0.415} &
		\multicolumn{1}{c}{0.880} &
		\multicolumn{1}{c}{1.962} &
		\multicolumn{1}{c}{$\chi^{2}$(1)=0.07, $p$=0.787} \\

		\multicolumn{1}{c}{Lucene} &
		\multicolumn{1}{c}{\shortstack{-0.364\textsuperscript{$\triangle$} \\ (0.20)}} &
		\multicolumn{1}{c}{\shortstack{0.238\textsuperscript{$\blacklozenge$} \\ (0.25)}} &
		\multicolumn{1}{c}{360.55} &
		\multicolumn{1}{c}{0.002} &
		\multicolumn{1}{c}{0.832} &
		\multicolumn{1}{c}{1.269} &
		\multicolumn{1}{c}{1.944} &
		\multicolumn{1}{c}{$\chi^{2}$(1)=0.86, $p$=0.353} \\

		\multicolumn{1}{c}{Hadoop} &
		\multicolumn{1}{c}{\shortstack{-1.992\textsuperscript{$\triangleleft$} \\ (0.61)}} &
		\multicolumn{1}{c}{\shortstack{0.893\textsuperscript{$\blacklozenge$} \\ (0.67)}} &
		\multicolumn{1}{c}{98.82} &
		\multicolumn{1}{c}{0.021} &
		\multicolumn{1}{c}{0.873} &
		\multicolumn{1}{c}{2.444} &
		\multicolumn{1}{c}{8.496} &
		\multicolumn{1}{c}{$\chi^{2}$(1)=2.00, $p$=0.157} \\

		\multicolumn{1}{c}{Ant} &
		\multicolumn{1}{c}{\shortstack{-1.504\textsuperscript{$\triangledown$} \\ (0.45)}} &
		\multicolumn{1}{c}{\shortstack{0.291\textsuperscript{$\blacklozenge$} \\ (0.51)}} &
		\multicolumn{1}{c}{138.64} &
		\multicolumn{1}{c}{0.002} &
		\multicolumn{1}{c}{0.595} &
		\multicolumn{1}{c}{1.337} &
		\multicolumn{1}{c}{3.274} &
		\multicolumn{1}{c}{$\chi^{2}$(1)=0.33, $p$=0.564} \\ 
		
		\hline
		
		\multicolumn{9}{l}{\tiny Significance of Coefficients (\textsf{R\textsuperscript{\textregistered}}): 0.001\textsuperscript{$\triangledown$}; 0.01\textsuperscript{$\triangleleft$}; 0.05\textsuperscript{$\triangleright$}; 0.1\textsuperscript{$\triangle$}; 1\textsuperscript{$\blacklozenge$}.
		\textsuperscript{$\otimes$}Significant Predictor: $p<\alpha$.} \\[-2.8ex]
	\end{tabular}
		
		
\end{table}

This research question aims at comparing the prevalence of clones between the classes that \smell{Complex Class} does not co-occur with \smell{Large Class} and those which these smells co-occur. The model to answer this research question has a predictor variable and this is a categorical variable that describes two categories of smell \smell{Complex Class}.  
Before building the logistic regression model, we performed a assumptions analysis of statistical test. In particular, we verify the possible frequency problems related to the dataset.
This analysis reveals that the dataset of all projects satisfy those requirements.

The next step of our analysis verifies the significance of the predictor variable (see \tablename~\ref{RQCCLCCC}). For this research question, the $p$-value of $\chi^2$ test of all projects is not less than 0.10. Thus, we can not reject the null hypothesis that the predictor variable and the prevalence of clones (dependent variable) are independent.
Therefore, for these software samples, there is no statistically significant evidence that the prevalence of clones in classes that exhibit only the smell \smell{Complex Class} is different from the prevalence of clones occurring in the classes where the smells \smell{Large Class} and \smell{Complex Class} co-occur.

%

\mbox{}\\
\noindent{\it \textbf{\ref{RQ-LC_LCCC}:} \RQLCLCCC~}

\begin{table}[H]
	\scriptsize
	\centering
	\def\arraystretch{2}
	\setlength{\tabcolsep}{2.2pt}
	\caption{Logistic Regression - \ref{RQ-LC_LCCC}.}
	\label{RQLCLCCC}
	\begin{tabular}{ccccccccc}
		\hline
		
		\multicolumn{1}{c}{\multirow{3}{*}{\textbf{Project}}} &
		\multicolumn{1}{c}{\multirow{3}{*}{\textbf{\shortstack{$\beta_0$ \\ (SE)}}}} &
		\multicolumn{1}{c}{\multirow{3}{*}{\textbf{\shortstack{$\beta_1$ \\ (SE)}}}} &
		\multicolumn{1}{c}{\multirow{3}{*}{\textbf{AIC}}} &
		\multicolumn{1}{c}{\multirow{3}{*}{\textbf{R\textsuperscript{2}}}} &
		\multicolumn{3}{c}{\textbf{IC for $\beta_1$}} &
		\multicolumn{1}{c}{\multirow{3}{*}{\textbf{\shortstack{\textit{Significance} \\ \textit{of} \\ \textit{Predictor}}}}} \\ \cline{6-8} 
		
		\multicolumn{1}{c}{}&
		\multicolumn{1}{c}{} &
		\multicolumn{1}{c}{} &
		\multicolumn{1}{c}{}&
		\multicolumn{1}{c}{}&
		\multicolumn{1}{c}{\multirow{2}{*}{\textbf{\shortstack{Lower \\ (5\%)}}}} &
		\multicolumn{1}{c}{\multirow{2}{*}{\textbf{OR}}} &
		\multicolumn{1}{c}{\multirow{2}{*}{\textbf{\shortstack{Upper \\ (95\%)}}}} &
		\multicolumn{1}{c}{} \\ 
		
		\multicolumn{1}{c}{}&
		\multicolumn{1}{c}{} &
		\multicolumn{1}{c}{} &
		\multicolumn{1}{c}{}&
		\multicolumn{1}{c}{}&
		\multicolumn{1}{c}{} &
		\multicolumn{1}{c}{} &
		\multicolumn{1}{c}{} &
		\multicolumn{1}{c}{} \\[-2.5ex] \hline
		
		
		\multicolumn{1}{c}{ArgoUML} &
		\multicolumn{1}{c}{\shortstack{-0.485\textsuperscript{$\triangle$} \\ (0.25)}} &
		\multicolumn{1}{c}{\shortstack{-0.195\textsuperscript{$\blacklozenge$} \\ (0.32)}} &
		\multicolumn{1}{c}{239.68} &
		\multicolumn{1}{c}{0.002} &
		\multicolumn{1}{c}{0.483} &
		\multicolumn{1}{c}{0.822} &
		\multicolumn{1}{c}{1.407} &
		\multicolumn{1}{c}{$\chi^{2}$(1)=0.36, $p$=0.548} \\

		\multicolumn{1}{c}{Cassandra} &
		\multicolumn{1}{c}{\shortstack{-1.887\textsuperscript{$\triangledown$} \\ (0.47)}} &
		\multicolumn{1}{c}{\shortstack{0.255\textsuperscript{$\blacklozenge$} \\ (0.54)}} &
		\multicolumn{1}{c}{131.64} &
		\multicolumn{1}{c}{0.002} &
		\multicolumn{1}{c}{0.550} &
		\multicolumn{1}{c}{1.291} &
		\multicolumn{1}{c}{3.387} &
		\multicolumn{1}{c}{$\chi^{2}$(1)=0.23, $p$=0.633} \\

		\multicolumn{1}{c}{Lucene} &
		\multicolumn{1}{c}{\shortstack{-0.452\textsuperscript{$\blacklozenge$} \\ (0.27)}} &
		\multicolumn{1}{c}{\shortstack{0.326\textsuperscript{$\blacklozenge$} \\ (0.32)}} &
		\multicolumn{1}{c}{297.35} &
		\multicolumn{1}{c}{0.004} &
		\multicolumn{1}{c}{0.821} &
		\multicolumn{1}{c}{1.386} &
		\multicolumn{1}{c}{2.367} &
		\multicolumn{1}{c}{$\chi^{2}$(1)=1.05, $p$=0.306} \\

		\multicolumn{1}{c}{Hadoop} &
		\multicolumn{1}{c}{\shortstack{-1.435\textsuperscript{$\triangleleft$} \\ (0.49)}} &
		\multicolumn{1}{c}{\shortstack{0.336\textsuperscript{$\blacklozenge$} \\ (0.57)}} &
		\multicolumn{1}{c}{105.93} &
		\multicolumn{1}{c}{0.004} &
		\multicolumn{1}{c}{0.568} &
		\multicolumn{1}{c}{1.400} &
		\multicolumn{1}{c}{3.807} &
		\multicolumn{1}{c}{$\chi^{2}$(1)=0.36, $p$=0.549} \\

		\multicolumn{1}{c}{Ant} &
		\multicolumn{1}{c}{\shortstack{$7.1E-15$\textsuperscript{$\blacklozenge$} \\ (0.63)}} &
		\multicolumn{1}{c}{\shortstack{-1.213\textsuperscript{$\triangle$} \\ (0.67)}} &
		\multicolumn{1}{c}{121.21} &
		\multicolumn{1}{c}{0.026} &
		\multicolumn{1}{c}{0.095} &
		\multicolumn{1}{c}{0.297} &
		\multicolumn{1}{c}{0.922} &
		\multicolumn{1}{c}{$\chi^{2}$(1)=3.09, $p$=0.079\textsuperscript{$\otimes$}} \\ 
		
		\hline

		\multicolumn{9}{l}{\tiny Significance of Coefficients (\textsf{R\textsuperscript{\textregistered}}): 0.001\textsuperscript{$\triangledown$}; 0.01\textsuperscript{$\triangleleft$}; 0.05\textsuperscript{$\triangleright$}; 0.1\textsuperscript{$\triangle$}; 1\textsuperscript{$\blacklozenge$}.
		\textsuperscript{$\otimes$}Significant Predictor: $p<\alpha$.} \\[-2.8ex]
	\end{tabular}
		

\end{table}

This research question is similar to the previous (\ref{RQ-CC_LCCC}), however, we investigate the prevalence of clones between the classes that \smell{Large Class} does not co-occur with \smell{Complex Class} and those which these smells co-occur. 

Based on our protocol of analysis, for each project, we observed that this dataset  has neither  assumption nor frequency problems (see Subsection \ref{AnalysisMethod}). 
Thus, 
we can build the logistic regression model.
Analyzing the significance of predictor of models (see \tablename~\ref{RQLCLCCC}), we observe that only for the Ant project the predictor variable and the prevalence of clones are associated. 
Thus, for other projects, the dependent and independent variables are not associated.

Considering only the logistic regression model of Ant project, we note that the coefficient $\beta_1$ also is significant and that the OR is less than one (0.29). This indicates that the prevalence of clones is more related to the isolated occurrence of smell \smell{Large Class} than the co-occurrence of smells \smell{Large Class} and \smell{Complex Class}. 
This means that when we compare the chances of clones between classes that have the isolated occurrence of smell \smell{Large Class} and those which the smells \smell{Large Class} and \smell{Complex Class} co-occur, the isolated occurrence of smell \smell{Large Class} increase in 29\% the chances of the class being involved with smell \smell{Duplicate Code}. However, for this case, only 2.6\% of the variability of the data can be explained by the model. Thus, this indicates that there are other factors that help to explain the prevalence of clones.


\global\def\TitleRQTwoX{The Intensity of Smells as a Factor}
\subsection{\TitleRQTwoX}
We investigate what happens to the prevalence of clones when the intensity of the smells \smell{Large} and/or \smell{Complex Classe} become more critical.

\mbox{}\\
\noindent{\it \textbf{\ref{RQ-CCIntensity_Clone}:} \RQCCIntensityClone}~

\begin{table}[H]
	\scriptsize
	\centering
	\def\arraystretch{2}
	\setlength{\tabcolsep}{2.2pt}
	\caption{Logistic Regression - \ref{RQ-CCIntensity_Clone}.}
	\label{TabRQ-CCIntensity_Clone}
	\begin{tabular}{ccccccccc}
		\hline

		\multicolumn{1}{c}{\multirow{3}{*}{\textbf{Project}}} &
		\multicolumn{1}{c}{\multirow{3}{*}{\textbf{\shortstack{$\beta_0$ \\ (SE)}}}} &
		\multicolumn{1}{c}{\multirow{3}{*}{\textbf{\shortstack{$\beta_1$ \\ (SE)}}}} &
		\multicolumn{1}{c}{\multirow{3}{*}{\textbf{AIC}}} &
		\multicolumn{1}{c}{\multirow{3}{*}{\textbf{R\textsuperscript{2}}}} &
		\multicolumn{3}{c}{\textbf{IC for $\beta_1$}} &
		\multicolumn{1}{c}{\multirow{3}{*}{\textbf{\shortstack{\textit{Significance} \\ \textit{of} \\ \textit{Predictor}}}}} \\ \cline{6-8} 
		
		\multicolumn{1}{c}{}&
		\multicolumn{1}{c}{} &
		\multicolumn{1}{c}{} &
		\multicolumn{1}{c}{}&
		\multicolumn{1}{c}{}&
		\multicolumn{1}{c}{\multirow{2}{*}{\textbf{\shortstack{Lower \\ (5\%)}}}} &
		\multicolumn{1}{c}{\multirow{2}{*}{\textbf{OR}}} &
		\multicolumn{1}{c}{\multirow{2}{*}{\textbf{\shortstack{Upper \\ (95\%)}}}} &
		\multicolumn{1}{c}{} \\ 
	
		\multicolumn{1}{c}{}&
		\multicolumn{1}{c}{} &
		\multicolumn{1}{c}{} &
		\multicolumn{1}{c}{}&
		\multicolumn{1}{c}{}&
		\multicolumn{1}{c}{} &
		\multicolumn{1}{c}{} &
		\multicolumn{1}{c}{} &
		\multicolumn{1}{c}{} \\[-2.5ex] \hline


		\multicolumn{1}{c}{ArgoUML} &
		\multicolumn{1}{c}{\shortstack{-0.810\textsuperscript{$\triangleright$} \\ (0.42)}} &
		\multicolumn{1}{c}{\shortstack{1.370\textsuperscript{$\triangleright$} \\ (0.75)}} &
		\multicolumn{1}{c}{50.51} &
		\multicolumn{1}{c}{0.069} &
		\multicolumn{1}{c}{1.166} &
		\multicolumn{1}{c}{3.937} &
		\multicolumn{1}{c}{14.512} &
		\multicolumn{1}{c}{$\chi^{2}$(1)=3.44, $p$=0.064\textsuperscript{$\otimes$}} \\

		\multicolumn{1}{c}{Cassandra} &
		\multicolumn{1}{c}{\shortstack{-1.609\textsuperscript{$\triangledown$} \\ (0.44)}} &
		\multicolumn{1}{c}{\shortstack{0.510\textsuperscript{$\blacklozenge$} \\ (0.93)}} &
		\multicolumn{1}{c}{45.44} &
		\multicolumn{1}{c}{0.007} &
		\multicolumn{1}{c}{0.307} &
		\multicolumn{1}{c}{1.666} &
		\multicolumn{1}{c}{7.251} &
		\multicolumn{1}{c}{$\chi^{2}$(1)=0.29, $p$=0.592} \\

		\multicolumn{1}{c}{Lucene} &
		\multicolumn{1}{c}{\shortstack{-0.465\textsuperscript{$\triangle$} \\ (0.24)}} &
		\multicolumn{1}{c}{\shortstack{0.331\textsuperscript{$\blacklozenge$} \\ (0.44)}} &
		\multicolumn{1}{c}{138.81} &
		\multicolumn{1}{c}{0.004} &
		\multicolumn{1}{c}{0.672} &
		\multicolumn{1}{c}{1.393} &
		\multicolumn{1}{c}{2.882} &
		\multicolumn{1}{c}{$\chi^{2}$(1)=0.57, $p$=0.452} \\ 
		
		
		
		\hline

		\multicolumn{9}{l}{\tiny Significance of Coefficients (\textsf{R\textsuperscript{\textregistered}}): 0.001\textsuperscript{$\triangledown$}; 0.01\textsuperscript{$\triangleleft$}; 0.05\textsuperscript{$\triangleright$}; 0.1\textsuperscript{$\triangle$}; 1\textsuperscript{$\blacklozenge$}.
		\textsuperscript{$\otimes$}Significant Predictor: $p<\alpha$.} \\[-2.8ex]
		
	\end{tabular}
\end{table}

This research question investigates the prevalence of clones between the classes where the smell \smell{Complex Class} occur at the low level of intensity and those that have this smell occurring at the high level of intensity. Observe that for this RQ, we do not consider the classes that have the co-occurrence of smells \smell{Large Class} and \smell{Complex Class}, which are investigated in  RQ2.3.

In order to answer this RQ, firstly we checked the assumptions of logistic regression.  Considering the contingency table of all projects, we observe that most part (77\%) of classes with smell \smell{Complex Class}\textsuperscript{$Low$} does not have clones. Calculating the expected frequencies, we observe that the Hadoop and Ant project do not satisfy the necessary conditions of the statistical test. Thus, these projects will not be analyzed.

\tablename~\ref{TabRQ-CCIntensity_Clone} denotes the output of logistic regression models for three projects. We observe that only for ArgoUML project, we reject the null hypothesis that the predictor variable and the prevalence of clones (dependent variable) are independent. Additionally, the significance of coefficient $\beta_1$ to the model of ArgoUML also is significant ($p-$value < $\alpha$). Thus, the value of Odds Ratio indicates the chances of the prevalence of clones to be associated with the intensity of smell \smell{Complex Class}. In particular, the OR denotes that the classes with smell \smell{Complex Class}\textsuperscript{$High$} have 3.93 times more chance to participate in clones than classes that have the smell \smell{Complex Class}\textsuperscript{$Low$}. However, for this case, only 6.9\% of the variability of the data can be explained by the model. This means that in addition to the intensity of this smell there are another factor(s) that help to explain the prevalence of clones.


\mbox{}\\
\noindent{\it \textbf{\ref{RQ-LCIntensity_Clone}:} \RQLCIntensityClone}~

Analogously to the \ref{RQ-CCIntensity_Clone}, we analyze the prevalence of clones into the class that have the smell \smell{Large Class}. Specifically, we investigate what happens with the prevalence of clones when the class have a low level of intensity of smell \smell{Large Class} or when the intensity of this smell is high.

The first step before building the logistic regression models is verify the numerical problems. Thus, we construct the contingency tables and calculate their expected frequencies. From contingency table, we observe that the most part (67\%) of classes with the smell \smell{Large Class}\textsuperscript{Low} does not have clones. Analyzing the expected frequencies, we observed that three projects (Cassandra, Hadoop and Ant) violate the assumptions related to the numerical problems (see Subsection \ref{AnalysisMethod}). Thus, these projects are not considered in the following analysis.

According to the \tablename~\ref{TabRQ-LCIntensity_Clone}, the chi-square test is not significant for any projects. In other words, we do not have any evidence that the predictor variable and prevalence of clones are dependent/related. Thus, considering the five projects, we can not found any evidence of the association between the prevalence of clones and the intensity of smell \smell{Large class}. Observe that this analysis does not consider the co-occurrence of smells \smell{Large class} with other smells. This will be considered on the next research question.

\begin{table}[H]
	\scriptsize
	\centering
	\def\arraystretch{2}
	\setlength{\tabcolsep}{2.2pt}
	\caption{Logistic Regression - \ref{RQ-LCIntensity_Clone}.}
	\label{TabRQ-LCIntensity_Clone}
	\begin{tabular}{ccccccccc}
		\hline
		
		\multicolumn{1}{c}{\multirow{3}{*}{\textbf{Project}}} &
		\multicolumn{1}{c}{\multirow{3}{*}{\textbf{\shortstack{$\beta_0$ \\ (SE)}}}} &
		\multicolumn{1}{c}{\multirow{3}{*}{\textbf{\shortstack{$\beta_1$ \\ (SE)}}}} &
		\multicolumn{1}{c}{\multirow{3}{*}{\textbf{AIC}}} &
		\multicolumn{1}{c}{\multirow{3}{*}{\textbf{R\textsuperscript{2}}}} &
		\multicolumn{3}{c}{\textbf{IC for $\beta_1$}} &
		\multicolumn{1}{c}{\multirow{3}{*}{\textbf{\shortstack{\textit{Significance} \\ \textit{of} \\ \textit{Predictor}}}}} \\ \cline{6-8} 
		
		\multicolumn{1}{c}{}&
		\multicolumn{1}{c}{} &
		\multicolumn{1}{c}{} &
		\multicolumn{1}{c}{}&
		\multicolumn{1}{c}{}&
		\multicolumn{1}{c}{\multirow{2}{*}{\textbf{\shortstack{Lower \\ (5\%)}}}} &
		\multicolumn{1}{c}{\multirow{2}{*}{\textbf{OR}}} &
		\multicolumn{1}{c}{\multirow{2}{*}{\textbf{\shortstack{Upper \\ (95\%)}}}} &
		\multicolumn{1}{c}{} \\ 
		
		\multicolumn{1}{c}{}&
		\multicolumn{1}{c}{} &
		\multicolumn{1}{c}{} &
		\multicolumn{1}{c}{}&
		\multicolumn{1}{c}{}&
		\multicolumn{1}{c}{} &
		\multicolumn{1}{c}{} &
		\multicolumn{1}{c}{} &
		\multicolumn{1}{c}{} \\[-2.5ex] \hline
		
		
		\multicolumn{1}{c}{ArgoUML} &
		\multicolumn{1}{c}{\shortstack{-0.405\textsuperscript{$\blacklozenge$} \\ (0.27)}} &
		\multicolumn{1}{c}{\shortstack{-0.693\textsuperscript{$\blacklozenge$} \\ (0.86)}} &
		\multicolumn{1}{c}{87.03} &
		\multicolumn{1}{c}{0.008} &
		\multicolumn{1}{c}{0.099} &
		\multicolumn{1}{c}{0.5} &
		\multicolumn{1}{c}{1.883} &
		\multicolumn{1}{c}{$\chi^{2}$(1)=0.70, $p$=0.402} \\


		\multicolumn{1}{c}{Lucene} &
		\multicolumn{1}{c}{\shortstack{-0.485\textsuperscript{$\blacklozenge$} \\ (0.31)}} &
		\multicolumn{1}{c}{\shortstack{0.149\textsuperscript{$\blacklozenge$} \\ (0.66)}} &
		\multicolumn{1}{c}{76.12} &
		\multicolumn{1}{c}{0.001} &
		\multicolumn{1}{c}{0.375} &
		\multicolumn{1}{c}{1.16} &
		\multicolumn{1}{c}{3.456} &
		\multicolumn{1}{c}{$\chi^{2}$(1)=0.05, $p$=0.823} \\



		\hline
		
		\multicolumn{9}{l}{\tiny Significance of Coefficients (\textsf{R\textsuperscript{\textregistered}}): 0.001\textsuperscript{$\triangledown$}; 0.01\textsuperscript{$\triangleleft$}; 0.05\textsuperscript{$\triangleright$}; 0.1\textsuperscript{$\triangle$}; 1\textsuperscript{$\blacklozenge$}.
		\textsuperscript{$\otimes$}Significant Predictor: $p<\alpha$.} \\[-2.8ex]
		
	\end{tabular}
		
		
		
\end{table}



\mbox{}\\
\noindent{\it \textbf{\ref{RQ-RQ_LCCCIntensity_Clone}:} \RQLCCCIntensityClone}~

This research question investigates the relation of the prevalence of clones with respect to the co-occurrence  of smells \smell{Large Class} and \smell{Complex Class}. In particular, we examined whether the intensity 
of these smells can be associated to the presence of clones. 
This means that we can build two models for each project, i) the first we considering the intensity of smell \smell{Large Class} and ii) the other taking into account the intensity of smell \smell{Complex Class}. In order to choose the better model we should evaluate each one of them.

As the first step, in order to obtain the expected frequencies, we construct the contingency table of each model for each project. For the model based on the intensity of smell \smell{Complex Class}, all projects satisfy the assumptions related to the numerical problems and we also observed that the most part of clones occurs when the class has the smell \smell{Complex Class} at the $High$ level of intensity. In other words, the models denoted by "\textsuperscript{$\star$}" in \tablename~\ref{TabRQ_LCCCIntensity_Clone}, should be evaluated.
We also had a similar result for the models based on the intensity of smell \smell{Large Class} (represented by "\textsuperscript{$\oslash$}" in \tablename~\ref{TabRQ_LCCCIntensity_Clone}). Thus, both models should be analyzed.

Next, we should choose the best model for our dataset. Thus we analyze the significance of predictors of all models. Whether two different models are significant for the same project, we use the Akaike Information Criterion (AIC) values. According to \citet{field2012discovering}, AIC is a measure of fit and can be used to deciding which of two models fits the data better. The lower the AIC, the better is the fit of the model. \tablename~\ref{TabRQ_LCCCIntensity_Clone} shows three projects with predictors significant (ArgoUML, Cassandra and Lucene) but we have only one situation where two different models are significant. This occurs with Lucene project and the analysis of AIC reveals that the best model is those based on the intensity of smell \smell{Complex Class}. Interesting to note that for ArgoUML and Cassandra, the best model also is those taking into account the intensity of smell \smell{Complex Class}. 
Thus, as resulting of this step, we conclude that the best model for all significant projects is based on  \smell{Complex Class} intensity (represented by "\textsuperscript{$\star$}" in \tablename~\ref{TabRQ_LCCCIntensity_Clone}) and the model denoted as "\textsuperscript{$\oslash$}" in \tablename~\ref{TabRQ_LCCCIntensity_Clone}, was not relevant for any project.

\begin{table}[H]
	\setlength{\arrayrulewidth}{0.2pt}
	\scriptsize
	\centering
	\def\arraystretch{2}
	\setlength{\tabcolsep}{2.2pt}
	\caption{Logistic Regression - \ref{RQ-RQ_LCCCIntensity_Clone}.}
	\label{TabRQ_LCCCIntensity_Clone}
	\begin{tabular}{cccccccccc}
		\hline
		
		\multicolumn{1}{c}{\multirow{3}{*}{\textbf{Project}}} &
		\multicolumn{1}{c}{}&
		\multicolumn{1}{c}{\multirow{3}{*}{\textbf{\shortstack{$\beta_0$ \\ (SE)}}}} &
		\multicolumn{1}{c}{\multirow{3}{*}{\textbf{\shortstack{$\beta_1$ \\ (SE)}}}} &
		\multicolumn{1}{c}{\multirow{3}{*}{\textbf{AIC}}} &
		\multicolumn{1}{c}{\multirow{3}{*}{\textbf{R\textsuperscript{2}}}} &
		\multicolumn{3}{c}{\textbf{IC for $\beta_1$}} &
		\multicolumn{1}{c}{\multirow{3}{*}{\textbf{\shortstack{\textit{Significance} \\ \textit{of} \\ \textit{Predictor}}}}} \\ \cline{7-9}
		
		\multicolumn{1}{c}{}&
		\multicolumn{1}{c}{}&
		\multicolumn{1}{c}{} &
		\multicolumn{1}{c}{} &
		\multicolumn{1}{c}{}&
		\multicolumn{1}{c}{}&
		\multicolumn{1}{c}{\multirow{2}{*}{\textbf{\shortstack{Lower \\ (5\%)}}}} &
		\multicolumn{1}{c}{\multirow{2}{*}{\textbf{OR}}} &
		\multicolumn{1}{c}{\multirow{2}{*}{\textbf{\shortstack{Upper \\ (95\%)}}}} &
		\multicolumn{1}{c}{} \\ 
		
		\multicolumn{1}{c}{}&
		\multicolumn{1}{c}{}&
		\multicolumn{1}{c}{} &
		\multicolumn{1}{c}{} &
		\multicolumn{1}{c}{}&
		\multicolumn{1}{c}{}&
		\multicolumn{1}{c}{} &
		\multicolumn{1}{c}{} &
		\multicolumn{1}{c}{} &
		\multicolumn{1}{c}{} \\[-2.5ex] \hline
		
		
		\multicolumn{1}{c}{\multirow{2}{*}{ArgoUML}} &
		\multicolumn{1}{c}{\textsuperscript{$\star$}} &
		\multicolumn{1}{c}{\shortstack{-1.178\textsuperscript{$\triangledown$} \\ (0.33)}} &
		\multicolumn{1}{c}{\shortstack{0.822\textsuperscript{$\triangleright$} \\ (0.41)}} &
		\multicolumn{1}{c}{151.79} &
		\multicolumn{1}{c}{0.027} &
		\multicolumn{1}{c}{1.169} &
		\multicolumn{1}{c}{2.275} &
		\multicolumn{1}{c}{4.561} &
		\multicolumn{1}{c}{$\chi^{2}$(1)=4.16, $p$=0.041\textsuperscript{$\otimes$}} \\ 
		
		\multicolumn{1}{c}{} &
		\textsuperscript{\textsuperscript{$\oslash$}} &
		\multicolumn{1}{c}{\shortstack{-0.356\textsuperscript{$\blacklozenge$} \\ (0.35)}} &
		\multicolumn{1}{c}{\shortstack{-0.462\textsuperscript{$\blacklozenge$} \\ (0.42)}} &
		\multicolumn{1}{c}{154.75} &
		\multicolumn{1}{c}{0.008} &
		\multicolumn{1}{c}{0.315} &
		\multicolumn{1}{c}{0.629} &
		\multicolumn{1}{c}{1.265} &
		\multicolumn{1}{c}{$\chi^{2}$(1)=1.20, $p$=0.274} \\

		\multicolumn{1}{c}{\multirow{2}{*}{Cassandra}} &
		\multicolumn{1}{c}{\textsuperscript{$\star$}} &
		\multicolumn{1}{c}{\shortstack{-2.970\textsuperscript{$\triangledown$} \\ (0.72)}} &
		\multicolumn{1}{c}{\shortstack{1.772\textsuperscript{$\triangleright$} \\ (0.77)}} &
		\multicolumn{1}{c}{94.715} &
		\multicolumn{1}{c}{0.075} &
		\multicolumn{1}{c}{1.886} &
		\multicolumn{1}{c}{5.886} &
		\multicolumn{1}{c}{26.881} &
		\multicolumn{1}{c}{$\chi^{2}$(1)=7.33, $p$=0.007\textsuperscript{$\otimes$}} \\ 

		\multicolumn{1}{c}{} &
		\textsuperscript{\textsuperscript{$\oslash$}} &
		\multicolumn{1}{c}{\shortstack{-2.197\textsuperscript{$\triangledown$} \\ (0.52)}} &
		\multicolumn{1}{c}{\shortstack{0.810\textsuperscript{$\blacklozenge$} \\ (0.60)}} &
		\multicolumn{1}{c}{100.06} &
		\multicolumn{1}{c}{0.02} &
		\multicolumn{1}{c}{0.877} &
		\multicolumn{1}{c}{2.25} &
		\multicolumn{1}{c}{6.663} &
		\multicolumn{1}{c}{$\chi^{2}$(1)=1.98, $p$=0.159} \\

		\multicolumn{1}{c}{\multirow{2}{*}{Lucene}} &
		\multicolumn{1}{c}{\textsuperscript{$\star$}} &
		\multicolumn{1}{c}{\shortstack{-1.321\textsuperscript{$\triangledown$} \\ (0.32)}} &
		\multicolumn{1}{c}{\shortstack{1.776\textsuperscript{$\triangledown$} \\ (0.38)}} &
		\multicolumn{1}{c}{200.28} &
		\multicolumn{1}{c}{0.113} &
		\multicolumn{1}{c}{3.206} &
		\multicolumn{1}{c}{5.906} &
		\multicolumn{1}{c}{11.35} &
		\multicolumn{1}{c}{$\chi^{2}$(1)=24.90, $p$=0.000\textsuperscript{$\otimes$}} \\ 		
		
		\multicolumn{1}{c}{} &
		\textsuperscript{\textsuperscript{$\oslash$}} &
		\multicolumn{1}{c}{\shortstack{-0.510\textsuperscript{$\triangleright$} \\ (0.25)}} &
		\multicolumn{1}{c}{\shortstack{0.636\textsuperscript{$\triangle$} \\ (0.32)}} &
		\multicolumn{1}{c}{221.39} &
		\multicolumn{1}{c}{0.017} &
		\multicolumn{1}{c}{1.103} &
		\multicolumn{1}{c}{1.888} &
		\multicolumn{1}{c}{3.267} &
		\multicolumn{1}{c}{$\chi^{2}$(1)=3.79, $p$=0.051\textsuperscript{$\otimes$}} \\

		\multicolumn{1}{c}{\multirow{2}{*}{Hadoop}} &
		\multicolumn{1}{c}{\textsuperscript{$\star$}} &
		\multicolumn{1}{c}{\shortstack{-1.658\textsuperscript{$\triangleleft$} \\ (0.54)}} &
		\multicolumn{1}{c}{\shortstack{0.822\textsuperscript{$\blacklozenge$} \\ (0.63)}} &
		\multicolumn{1}{c}{78.686} &
		\multicolumn{1}{c}{0.023} &
		\multicolumn{1}{c}{0.834} &
		\multicolumn{1}{c}{2.275} &
		\multicolumn{1}{c}{7.043} &
		\multicolumn{1}{c}{$\chi^{2}$(1)=1.79, $p$=0.181} \\ 		
		
		\multicolumn{1}{c}{} &
		\textsuperscript{\textsuperscript{$\oslash$}} &
		\multicolumn{1}{c}{\shortstack{-1.609\textsuperscript{$\triangleleft$} \\ (0.54)}} &
		\multicolumn{1}{c}{\shortstack{0.740\textsuperscript{$\blacklozenge$} \\ (0.63)}} &
		\multicolumn{1}{c}{79.04} &
		\multicolumn{1}{c}{0.019} &
		\multicolumn{1}{c}{0.767} &
		\multicolumn{1}{c}{2.096} &
		\multicolumn{1}{c}{6.499} &
		\multicolumn{1}{c}{$\chi^{2}$(1)=1.44, $p$=0.230} \\

		\multicolumn{1}{c}{\multirow{2}{*}{Ant}} &
		\multicolumn{1}{c}{\textsuperscript{$\star$}} &
		\multicolumn{1}{c}{\shortstack{-1.312\textsuperscript{$\triangleleft$} \\ (0.42)}} &
		\multicolumn{1}{c}{\shortstack{0.149\textsuperscript{$\blacklozenge$} \\ (0.51)}} &
		\multicolumn{1}{c}{107.26} &
		\multicolumn{1}{c}{0.001} &
		\multicolumn{1}{c}{0.503} &
		\multicolumn{1}{c}{1.16} &
		\multicolumn{1}{c}{2.814} &
		\multicolumn{1}{c}{$\chi^{2}$(1)=0.08, $p$=0.773} \\ 		
		
		\multicolumn{1}{c}{} &
		\textsuperscript{\textsuperscript{$\oslash$}} &
		\multicolumn{1}{c}{\shortstack{-1.550\textsuperscript{$\triangledown$} \\ (0.41)}} &
		\multicolumn{1}{c}{\shortstack{0.545\textsuperscript{$\blacklozenge$} \\ (0.51)}} &
		\multicolumn{1}{c}{106.18} &
		\multicolumn{1}{c}{0.011} &
		\multicolumn{1}{c}{0.755} &
		\multicolumn{1}{c}{1.72} &
		\multicolumn{1}{c}{4.158} &
		\multicolumn{1}{c}{$\chi^{2}$(1)=1.16, $p$=0.281} \\
		
		\hline
		
		\multicolumn{10}{l}{\tiny Significance of Coefficients (\textsf{R\textsuperscript{\textregistered}}): 0.001\textsuperscript{$\triangledown$}; 0.01\textsuperscript{$\triangleleft$}; 0.05\textsuperscript{$\triangleright$}; 0.1\textsuperscript{$\triangle$}; 1\textsuperscript{$\blacklozenge$}.
		\textsuperscript{$\otimes$}Significant Predictor: $p<\alpha$.} \\[-2.8ex]
		\multicolumn{10}{l}{\tiny \textsuperscript{$\oslash$} Model considering the intensity of the smell \textsc{Large Class} ($Y^{\prime})$.}\\[-2.8ex]
		\multicolumn{10}{l}{\tiny \textsuperscript{$\star$} Model based on the intensity of the smell \textsc{Complex Class} ($Y^{\prime\prime}$).}\\[-2.8ex]
	\end{tabular}
		
		

		

\end{table}

Therefore,  we take into account only the projects ArgoUML, Cassandra, Lucene, and their models marked with "\textsuperscript{$\star$}" in \tablename~\ref{TabRQ_LCCCIntensity_Clone}. Odds Ratio varies between 2.2 and 5.9, indicating that when smells \smell{Large Class} and \smell{Complex Class} co-occur in a class, the chances  of that class has clone is 2.2---5.9 times larger in classes that has \smell{Complex Class}\textsuperscript{$High$} than classes with \smell{Complex Class}\textsuperscript{$Low$}. However, only 2.7\%---11.3\% of the variability of the data can be explained by the model, indicating that there are other factors that help to explain the prevalence of clones and these considerations are consistent with those presented in the previous research questions


\section{Discussion} \label{Discussion}

This section provides a qualitative perspective of our results. In particular, we discuss the  statistically significant models in terms of their coefficients and predictors. For each one of these models, we examine the classes of projects that exhibit fragments of clones and have at least one of the smells of interest (\smell{Large Class} and/or \smell{Complex Class}). For instance, in the qualitative analysis of \ref{RQ-CCIntensity_Clone}, we inspect the Java classes of ArgoUML that have clones and the smell \smell{Large Class}. In this case, we inspect only ArgoUML project because the model of this project is the only statistically significant.

\subsection{\TitleRQOneX}
This part of the discussion is related to the analysis of smells considering only their frequencies, namely \ref{RQ-CC_LCCC} and \ref{RQ-LC_LCCC}.

\noindent\textbf{\ref{RQ-CC_LCCC}}. We studied the prevalence of clones in classes classified as \smell{Complex Class} and they are organized into two groups according to their smells (only CC and co-occurrence of LC/CC). The results, for all projects, showed that the single occurrence of smell CC is not significant to the prevalence of clones. This observation is also valid for classes that have the co-occurrence of smells LC and CC. Moreover, the frequency of classes with co-occurrences of LC/CC is more than twice ($\times 2.31$), in average, the occurrence of CC only, indicating that although \smell{CC} tends also to be \smell{LC}, the fact of being large does not increase their chance to have clones. 

%

\noindent\textbf{\ref{RQ-LC_LCCC}}. 
Its similar to the previous but the smell \smell{Large Class} is used to organize the classes into groups. 
Statistical analysis has shown that the prevalence of clones is more associated to the single occurrence of smell \smell{Large Class} than the co-occurrence of \smell{Large \& Complex Class} ($\beta_1=-1,213, OR=0,29$).
However, this observation is valid only for Ant project and we observe that classes with LC and CC are almost the triple ($\times 2.9$ in average) than LC-only classes. So, we observe that the fact of being large 
does not increase their chance to have clones, and in Ant was the inverse. Thus, the results are project-specific. Moreover, to understand why Ant had LC-only classes with higher chances to have clones than LC-CC classes, we performed a qualitative analysis in the 10 classes classified as LC-only, where  50\% of them also have the smell \smell{Duplicate Code}. From these classes with clones, we observed that the 80\% are "derivations" of the same entity, e.g., \textsf{CCMklbtype} has defined different attributes from those defined inside the class \textsf{CCMkattr}, but these classes shares: i) fully cloned methods (e.g., \textsf{getCommentCommand, getVersionCommand}); ii) partially cloned methods (e.g., \textsf{execute, checkOptions}) and iii) other methods (ex. \textsf{getVOB, getTypeValueCommand}).
This is an indication that these clones are practically the same, and they are distributed across several entities, suggesting that these classes provide little variability of data.

%
These clones could be avoided if adequate object-oriented practices (design patterns) had been adopted within those entities. 
Analyzing commits in these classes, we have observed that changes applied in one of these classes are commonly applied to others, where adaptations are minimal, reinforcing that object-oriented principles were not adequately followed. In the end, this finding could be helpful to indicate that when the system has high number of larges classes with high prevalence of clones those classes would deserve special attention.

\subsection{\TitleRQTwoX} 

This subsection  discuss qualitatively the results on the association of prevalence clones and the smells intensity of \smell{Large Class} and \smell{Complex Class} (\ref{RQ-CCIntensity_Clone}, \ref{RQ-LCIntensity_Clone} and \ref{RQ-RQ_LCCCIntensity_Clone}).


\noindent\textbf{\ref{RQ-CCIntensity_Clone}}. The statistical results revealed that from the three  analyzed projects (ArgoUML, Cassandra and Lucene), only one (ArgoUML) provide evidence for the association between the prevalence of clones and the intensity of \smell{Complex Class}. In this case, we observed that classes with \smell{Complex Class}\textsuperscript{$High$} are more clone-prone than other smelly classes affected by \smell{Complex Class}\textsuperscript{$Low$}. Specifically, the measure OR reveals that the chances are 3.93 times bigger. For this project, we have 15 classes with clones, where eight are classified as \smell{Complex Class}\textsuperscript{Low} and the rest as \smell{Complex Class}\textsuperscript{High}. We examined these 15 classes, and  report the main observations.

\begin{table}[!htbp]
	\noindent
	\scriptsize
	\def\arraystretch{1.0}
	\begin{tabular}{p{0.9\linewidth}}
		\begin{lstlisting}[firstline=1,lastline=131]{}
		(*$\dots$*)
		@Override
		public boolean stillValid(ToDoItem i, Designer dsgr) {
			if (!isActive()) {
				return false;
			}
			ListSet offs = i.getOffenders();
			Object f = offs.get(0);
			if (!predicate(f, dsgr)) {
				return false;
			}
			ListSet newOffs = computeOffenders(f);
			boolean res = offs.equals(newOffs);
			return res;
		}
		\end{lstlisting}	
		\\[-2.8ex]
	\end{tabular}
	\captionof{JavaCode}{\scriptsize Clone with low complexity (\textit{\textsf{CrUnconventionalAttrName}} -- ArgoUML).}
	\vspace{-0.5cm}
	\label{CloneCrUnconventionalAttrName}
\end{table}

Analyzing clone fragments from CC\textsuperscript{$Low$} classes, most of them (91\%) are simple and/or small clones. For example,  the class \textsf{CrUnconventionalAttrName} have three simple clone fragments, and one of them is partially in \codigo~\ref{CloneCrUnconventionalAttrName}. This clone refers to a low complexity functionality. For clones CC\textsuperscript{$Low$}, we still observe that the used code template may influence the size of clones. Other examples 
exception handling and/or undesired condition with similar handling. This is highly prevalent in clones having nested \textit{throw/try/catch}. 


On the other side, clone fragments of CC\textsuperscript{$High$} classes have shape, size and/or complexity different from those in CC\textsuperscript{$Low$} classes. 
In CC\textsuperscript{$High$} classes, we found no clones with complete functionality as the one in \codigo~\ref{CloneCrUnconventionalAttrName}. In these classes clones are smaller fragments in a large sized functionality. For example, the method \textsf{parseAttribute} in the class \textsf{AttributeNotationUml} has 229 lines (\smell{Long Method}),  and has internally two different clones. 
Moreover, clones in CC\textsuperscript{$High$} classes also correlate with the class itself having more control flow structures, i.e., are more complex too.

The average size of clones in CC\textsuperscript{$High$} classes is larger compared to size of clones in CC\textsuperscript{$Low$}, and they are concentrated in the longer methods.


From the qualitative analysis, we conclude that for ArgoUML, the high complexity of a class is associated with the shape, size and complexity of the clones, corroborating with the statistical result.

\noindent\textbf{\ref{RQ-LCIntensity_Clone}}. This analysis is focused only on LC classes which are not also CC. Statistical analysis has shown that the intensity of LC in these classes is  not a relevant factor to explain the prevalence of clones. For all systems, there are just a few  (12.6\%) LC\textsuperscript{$High$} classes, and most of them are LC\textsuperscript{$Low$}. Because one of the classes is highly dominant against the other, no significant effect could be observed. 

 \noindent\textbf{\ref{RQ-RQ_LCCCIntensity_Clone}}. In this question, we consider only entities where LC and CC co-occur. Statistical models were significant for ArgoUML, Cassandra and Lucene, which are used in this discussion. Odds ratio indicated chance of clones being found in CC\textsuperscript{$High$} to be 2.2-5.9 higher than being found in CC\textsuperscript{$Low$}.

In this part, we investigated how those clones found in ArgoUML, Cassandra and Lucene, could be classified over the type intra-class, inter-class ou mixed (both intra and inter).  
We manually analyzed each clone fragment of classes that manifest one or more LC/CC. 

In ArgoUML, from classes with clones, 15 classes have CC-only, 24 have LC-only and 40 have both CC and LC co-occurring.
The clones of classes CC-only or LC-only are almost all inter-class clones (95.8\%). The clones of classes where CC and LC co-occur have also a high rate of inter-class clones (72.3\%). We also analyzed Cassandra clones, which were similar to ArgoUML.
In Lucene, inter-class clones in classes CC-only or LC-only correspond to 70.9\% of clones and clones of classes where CC and LC co-occur have an even smaller rate of inter-class clones (36.2\%), indicating that there are inherent project-specific design factors that induce the type of cloning (inter or intra) developers adopt.

We have observed that mixed clones (intra and inter class) have a very low frequency.
The analysis of this RQ has shown  that (i) clones occurring in entities having smells CC and LC co-occurring tend to be localized in  entities with that same characteristic;  (ii) the isolated occurrence of LC or CC does not prevent inter-class clones; (iii) the typical type of clone (intra or inter-class) seems to be related to project-specific decisions. These observations are valid only for systems (ArgoUML, Cassandra, Lucene) where the association is significant.

\subsection{Practical Implications} \label{PracticalImplications}

The results for \ref{RQ-RQ_LCCCIntensity_Clone} have shown that clone fragments of large and complex classes are mostly restricted to that kind of class, i.e., clones occurring in classes where smells LC and CC co-occur are typical clones confined in other classes where  smells LC and CC also co-occur. In other words, it is not likely that clones occur, at the same time, in large and complex classes, and also in a class that is not large and complex. This finding could be used to optimize the detection of \smell{Duplicate Code}, that is, the detection algorithm could prune the search space for finding clones, in the sense, that large and complex class could be clustered to define a search space to look for clones.

Another practical implication is related to refactoring planning. As we have discussed, a reasonable part of clone fragments that occur in classes with just one of the smells (CC or LC) are inter-class clones. Also, inter-class clones are also highly prevalent in CC-LC classes, except for Lucene (36.2\%). The implication is that removing this kind of clone would require a more sophisticated operation, e.g., the application of \textit{Extract Superclass} or \textit{Extract Class}, and thus making this process more risky and laborious.
On the other hand,  intra-class clones would be 
simpler to be removed applying the \textit{Extract Method} refactoring. 
So, this kind of refactoring, in general, would be easier to be applied because it is local to just one class, i.e., it does not require investigating multiple classes. The implication is that this kind of refactoring could be applied more quickly, and could also be assigned for less experienced members of the development team, or those members of the team that do not have a broader knowledge of the system. 
However, this is just a hypothesis that should be further confirmed by other empirical study with human subjects.


\section{Threats to Validity} \label{ThreatsValidity}

This section discusses the threats that could affect the validity of our study. 
Some threats should be considered in the analysis of the presented results. In this context, the internal validity examines whether the findings of an investigation are aligned with the study population. 
On the other hand, external validity refers to the extent to which research results can be generalized to other conditions.

One of the threats to external validity occurs because our results can not be generalized to other  object-oriented programming languages, it because all the analyzed projects were developed in Java. To generalize this study, it is necessary to evaluate projects from other languages that also use the object-oriented paradigm. Furthermore, our empirical study only considers \textit{open source} projects and the most part of them (80\%) are maintained by \textit{Apache Foundation}. 
Another threat is related to the sample size (five projects of software). However, the sample of three or more software is usual in exploratory studies. This type of study is characterized by the production of preliminar evidence that would support more comprehensive and extensive studies.

With respect to the internal validity, the analysis  in this study does not include test classes  (e.g., classes related to the {\it JUnit} test cases). These classes have different characteristics from those used in production (e.g., a preliminary analysis of our data revealed that they have larger clones in terms of LOC). We think that developers and/or researchers are more interested in the occurrence of smells on the production classes than the occurrence of smells in test classes. For classes of test cases, we found specifics studies \citep{PA302, P189}. In general, our sample is aligned with the interests of the community.

The process of detecting the smells also is a threat to internal validity. According to \citet{deMello:2017}  performing ad-hoc manual identification of smells does not assure more effective results, so manually validating the detected smells would still not eliminate the threat. They observed that different context factors may influence on the conclusion about the incidence of a code smell. These factors are addressed to human aspects, such as the interaction among individuals and their professional roles. 
Moreover, DECOR is a state-of-the-art smell detector having a high accuracy in detecting smells, according to \citet{P154} this tool has up to 88\% of \textit{Precision} and 100\% of \textit{Recall}. PMD also is good to detecting identical clones \citep{Roy2009a}.

Finally, another threat to the internal validity is related to the scope and type of clones. This study considers only the clones of type I that occurred in smelly classes (\smell{Large Class} and/or \smell{Complex Class}). Nevertheless, there are clones in classes that have not been classified with these smells and these clones may present in a totally different way from those that we analyzed.
However, the focus of this study is investigate the interaction of clones of type I and the smells LC and/or CC. Therefore, the investigation of other types of clones and/or the possible occurrence of them with other smells or classes without smells should be carried out in future works.

\section{Conclusion} \label{Conclusion}

We investigated if the occurrence of smells \smell{Large Class} and \smell{Complex Class} could impact on the occurrence of clones. We considered different ways on how clones and these smells could interact: clones occurring in large-only classes, clones occurring in complex-only classes and clones occurring in classes that are simultaneously large and complex. Moreover, we investigate if the intensity of the LC and/or CC smells also plays a role in the prevalence of clones.

Our results have shown that conclusions are project-specific, i.e., they are not valid for all studied systems.
In particular, only for Ant that the prevalence of clones is  more associated to the co-occurrence of the smell \smell{Large Class} and \smell{Complex Class} than to the isolated occurrences of 
\smell{Large Class} (see \tablename~\ref{RQLCLCCC}). So, the general hypothesis of prevalence of clones being associated with co-occurrence of LC and CC does not hold.

However, as we consider the intensity of smells in entities that have both the smells \smell{Large Class} (LC) and \smell{Complex Class} (CC), our data indicates that clone prevalence may be associated to the $High$ intensity level of smell CC, either in CC only classes (ArgoUML), or CC co-occurring with LC (ArgoUML, Cassandra, Lucene), which in fact, may suggest a difference conclusion compared to that of \citet{fowler1999refactoring} quote that ``\textit{...a class with too much code is prime breeding ground for duplicated code...}". We would prefer to say that a class with highly complex code, and with too much code, would be prime breeding ground for duplicated code. 
On the other side, independently on how our studies on smells have been carried out, they explain only a small part of clone occurrence, i.e., there are other factors that would help to explain the occurrence of clones in LC and/or CC classes. 

We also observed that the class complexity and the intensity of the LC and CC smells influence some characteristics of the clone: shape (complete functionality or partial functionality clone), type (intra-inter clone), locality (clones restricted only to classes with specific configuration of smells), and size (LOC) of clones. 


Future work includes i) replicating  our study on proprietary systems, ii) enlarge the sample size, iii) investigate the influence of other smells on the prevalence of clones and iv) empirically investigate the practical implications.

\bibliographystyle{ACM-Reference-Format}
\balance
\bibliography{BibPapersRef}

\end{document}